\documentclass[aps,twocolumn,floatfix,footinbib]{revtex4-1}
\usepackage{epsfig}
\usepackage{graphicx}
\usepackage{dcolumn}
\usepackage{bm}
\usepackage{mathrsfs}
\usepackage{amsmath}
\usepackage{bbold}
\usepackage{color}
\usepackage{epstopdf}

\usepackage[colorlinks=true,pdfborder=001,linkcolor=blue,anchorcolor=blue,citecolor=blue,urlcolor=blue]{hyperref}

\begin{document}
\title{Theory of Fano resonance in single molecule electroluminescence induced by a scanning tunneling microscope}
\author{Lei-Lei Nian}
\author{Jing-Tao L\"u}
\affiliation{School of Physics and Wuhan National High Magnetic Field Center, Huazhong University of Science and Technology, 430074 Wuhan, P. R. China}

\begin{abstract}
The coupling between molecular exciton and gap plasmons plays a key role in
single molecular electroluminescence induced by a scanning tunneling microscope
(STM). But it has been difficult to clarify the complex experimental phenomena.
By employing the nonequilibrium Green's function method, we propose a general
theoretical model to understand the light emission spectrum from single molecule and gap
plasmons from an energy transport point of view.
The coherent interaction between gap plasmons and molecular exciton leads to
a prominent Fano resonance in the emission spectrum. We analyze the dependence of the Fano
line shape on the system parameters, based on which we provide a unified
account of several recent experimental observations. Moreover, we highlight the effect of the tip-molecule electronic coupling on the spectrum, which has hitherto not been considered.
\end{abstract}

\maketitle

\section{Introduction}

Recently, single molecular electroluminescence (EL) induced by the inelastic
electron tunneling from a scanning tunneling microscope (STM) has attracted a
lot of attention, yielding many fascinating physics and potential
applications\cite{qiu2003vibrationally,dong2004vibrationally,dong2010generation,schneider2012plasmonic,galperin2017photonics,kuhnke2017atomic,wang2016molecular,heiien1986photon}.
In such STM-induced luminescence (STML) experiments, light emission from gap
plasmon modes is a common
process\cite{berndt1993photon,gimzewski1999nanoscale,hoffmann2002tunneling,schull_electron-plasmon_2009,schneider_optical_2010,schneider2012light,lu_light_2013},
which in turn can dominate, accompany, or influence the luminescence of single
molecules positioned nearby STM
tip\cite{dong2010generation,liu2007bias,schneider2012plasmonic,osley2013fano}.
The resulting coupling between the molecular exciton and gap plasmons is of
interest because it contributes to the study of fundamental quantum phenomena,
including coherent energy transfer, cavity quantum electrodynamics, and
entanglement\cite{tame2013quantum}.

The importance of coherent interaction between molecular exciton and gap
plasmons in STML has been revealed in recent
experiments\cite{nordlander2014molecular,tan2014quantum,benz2014nanooptics,imada2016real,du2016chip,lerch2016quantum,yoshioka2016real,chikkaraddy2016single,li2016transformation,
nijs2017plasmonic,zhang2017sub,imada2017single,doppagne2017vibronic,kroger2018fano},  through,
for example, prominent Fano\cite{fano1961effects} line shapes in the emission spectrum.  Its possible
applications in single molecule detection\cite{zhang2017sub,imada2017single,doppagne2017vibronic}, single photon generation\cite{zhang2017electrically} have been
envisioned. Although these experiments show the important role played by the
coherent interaction between molecular exciton and gap plasmons, 
a systematic theoretical model to account for all these experimental results is so far lacking.
Revealing the connection between the line shape and the system parameters is important
for further development and application of this technique.

Here, we propose a general theoretical model that is able to account for all these experimental results.
We first demonstrate the coherent optical coupling between molecular exciton
and gap plasmons leads to a pronounced Fano resonance, whose line shape depends
sensitively on the system parameters.  Using
experimentally based parameters, the simulated spectrum shows quantitative
agreement with the experimental results. This is an essential step to predict or control the dynamic
energy transfer process in STML experiments.  

\begin{center}
\begin{figure*}
\includegraphics[scale=0.35]{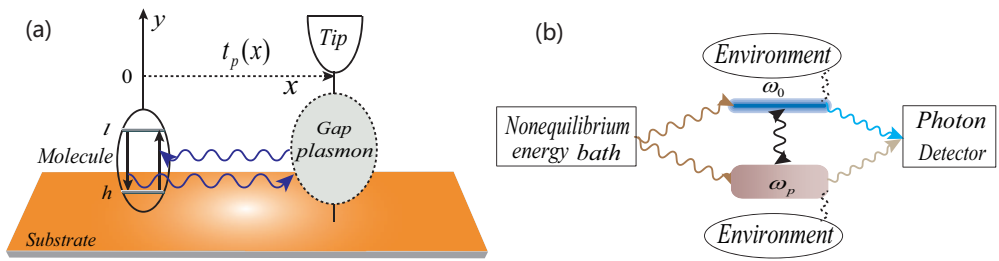}
\caption{(a) Schematic picture of the experimental setup in STM induced light emission and single molecule luminescence.
(b) Effective model to study the energy transfer in the experimental setup in (a). The nonequilibrium electronic bath includes the STM tip, the substrate and the single molecule under certain voltage bias. The gap plasmon is represented by a photon mode with angular frequency $\omega_p$, while the molecular exciton by a mode with frequency $\omega_0$. There is a direct coupling whose magnitude depends on the relative position of the STM tip and the molecule $x$ (see (a)). The emitted light is collected by the photon detector. There are also non-radiative channels into which the two photon modes can dissipate their energy, represented by the environment.}\label{Fig01}
\end{figure*}
\end{center}

\begin{center}
\begin{figure}[h]
\includegraphics[scale=1.2]{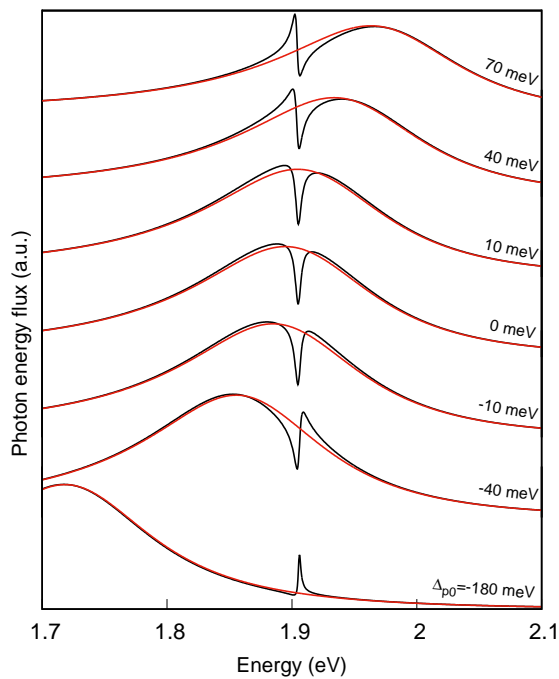}
\caption{The energy detuning $\Delta_{p0}$ dependence of flux with $t_{p}(x)=0.013$~eV, the red lines represent the spectrum of gap plasmon mode. The other parameters are the same as in Fig. \ref{Fig02} (II), and keep fixed. 
}\label{Fig03}
\end{figure}
\end{center}

\begin{center}
\begin{figure*}[t]
\includegraphics[scale=0.39]{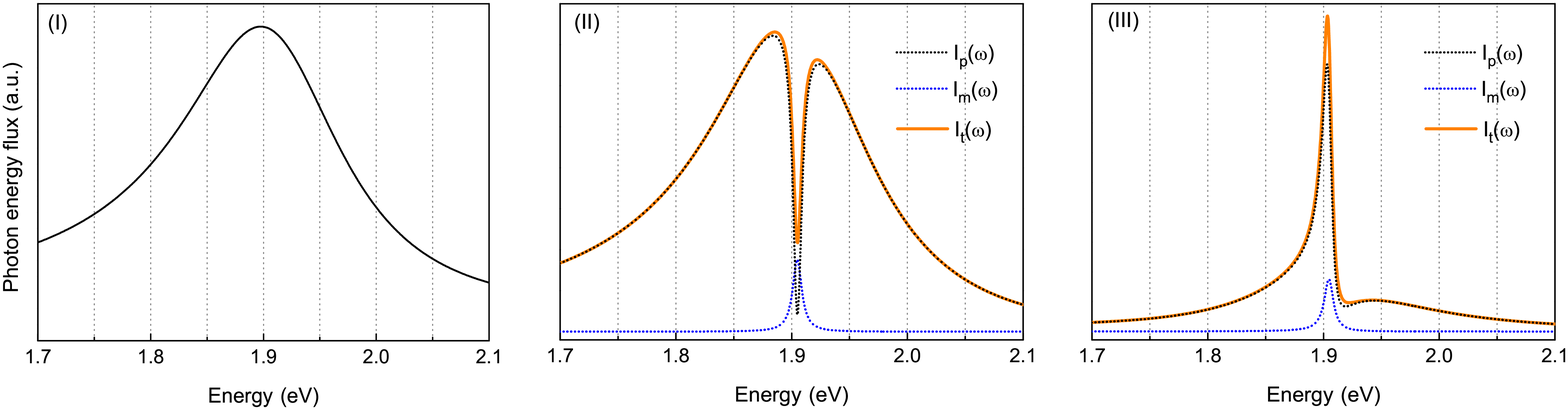} 
\caption{ The photon energy flux calculated as the tip is: (I) apart from the molecule, (II) slight aside from the molecule so that there is a coupling between gap plasmon and the molecular exciton, but no direct electronic coupling between the tip and the molecule, (III)  located on top of the molecule, so that there is direct electronic coupling between the tip and the molecule. The separate contributions from the plasmon and the molecular exciton are shown as dotted lines, while the sum of the two is shown as red solid line.
}
\label{Fig02}
\end{figure*}
\end{center}

\section{Model and Theory}
We consider a model system schematically shown in Fig.~\ref{Fig01} (a). 
The voltage bias applied between the tip and the substrate generates a flowing electrical current between them, which is used to 
excite the localized gap plasmons. 
A single  molecule is represented by two electronic states $l$ and $h$, 
representing the lowest unoccupied molecular orbital (LUMO) and highest occupied molecular orbital (HOMO), 
respectively.  
If the molecule is present underneath the STM tip, under certain bias, a molecular exciton may also be created by the electrical current, i.e., injecting an electron to the LUMO and a hole to the HOMO orbital. If the molecule is not far away from the tip, the molecular exciton can be created by the gap plasmons, given its much wider frequency and larger spatial distribution. 
This requires a direct coupling between the  gap plasmons and the molecular exciton.

We model the gap plasmon using a photon field with angular frequency $\omega_p$. In reality, there could
be several modes with similar frequencies. Similarly, we model the molecular exciton using a photon field with angular frequency $\omega_0$. The two photon fields couple to each other through the parameter $t_p(x)$, which depends on the tip-molecule distance $x$. 

To study the energy transfer between the electron and photon fields, we use an effective model shown in Fig.~\ref{Fig01} (b). 
The biased electronic system acts as an effective nonequilibrium energy bath, which supplies energy to the gap plasmons and the molecular exciton.
The energy absorbed by the photon fields is either dissipated into the environment or radiated to the free space. The radiation
then goes to the detector. 

Energy transport for this effective model can be studied using the nonequilibrium Green's function (NEGF) method\cite{meir1992landauer,wang2008quantum,haug2008quantum,mahan2013many}. 
The frequency-resolved energy flux going into bath $\alpha$ is written as
\begin{equation}\label{1}
I_{\alpha}(\omega)=\frac{\omega}{2\pi} {\rm Tr}\left[\Pi^{<}_{\alpha}(\omega)D^{>}(\omega)-\Pi^{>}_{\alpha}(\omega)D^{<}(\omega)\right].
\end{equation}
Here,  $D^>$ ($D^<$) is the greater (less) Green's function of the photon fields, $\Pi^>_{\alpha}$ ($\Pi^<_{\alpha}$) is the corresponding self-energy due to coupling to bath $\alpha$. We have considered three kinds of baths: (1) the nonequilibrium electronic system which supplies the energy, thus $I_{el}<0$; (2) the photon detector which collects the radiation and corresponds to the measured photon flux; (3) the non-radiative environment into which the non-radiative energy goes. The two terms in Eq.~(\ref{1}) correspond to energy flowing into and out of the bath, respectively. The Green's functions and self-energies in Eq.~(\ref{1}) are solved within the self-consistent Born approximation (SCBA)\cite{galperin2004line,frederiksen2004inelastic,haug2008quantum,lu2007coupled}. The photon flux can then be calculated from them. The details of the method can be found in the Appendices \ref{A} and \ref{B}.

\section{Results and Discussions}
\subsection{Important parameters}
The advantage of the effective model is that, we separate the electronic part of the whole system from the photonic part. All the electronic part, including the STM tip, molecule and substrate, is modeled as a nonequilibrium energy bath. The most important feature of the nonequilibrium bath is that, the width of its energy spectrum is determined by the applied bias $|eV|$, i.e., the bath can not excite photon mode whose energy is larger than the applied bias $\omega>|eV|$. It enters into our theory through the self-energy $\Pi_{el}$, on which the Green's functions $D^>$ and $D^<$ in Eq.~(\ref{1}) depend. Meanwhile, the line shape of the spectrum is mainly determined by the two parameters describing the photon modes, which we consider in the following.

The first important parameter that determines the line shape is the detuning $\Delta_{p0}  = \omega_p - \omega_0$.
In the experiment, the resonant frequency of gap plasmon $\omega_{p}$ can be tuned by adjusting the tip shape, or modifying the dielectric properties of the substrate, i.e., introducing dielectric layers. Figure~\ref{Fig03} displays the evolution of the spectrum with different values of energy detuning $\Delta_{p0}$.
We note that the strong energy detuning dependence of the Fano line shape is in agreement with experimental findings \cite{zhang2017sub,imada2017single,chong2016narrow} and can be fitted by a simple model detailed in Appendix \ref{C} [Eq.~(\ref{C11}) or (\ref{C12})], where the magnitude of the Fano $q$ factor is mainly determined by the detuning $q \propto -\Delta_{p0}$.

The second parameter is the coupling between the gap plasmon and the molecular exciton $t_p(x)$, 
determined by the relative position of the tip and the molecule ($x$),
according to which, we can define three regimes.
They correspond to the tip apart from (I), slight aside from (II) and  located on top of the molecule (III), respectively.
The energy flux spectrum of different situations is plotted in Fig.~\ref{Fig02}\footnote{In the calculations, we set $t_{1l,h}=t_{ma1}$, $t_{2l,h}=t_{ma2}$, $\Gamma_{a1,2}=\Gamma_{ae}$ and
$\Gamma_{ls}=\Gamma_{hs}=\Gamma_{ms}$. The parameters producing the plots are the following: (I) $\Gamma_{ms}=0.5$~eV, $t_{ma1}=0$, $t_{ma2}=0$, $\gamma_{d0}=0$, $\gamma_{0e}=0$ and $t_{p}(x)=0$; (II) $\Gamma_{ms}=0.5$~eV,  $t_{ma1}=0$, $t_{ma2}=0$, $\gamma_{d0}=2.5\times10^{-4}$~eV, $\gamma_{0e}=1.5\times10^{-3}$~eV and $t_{p}(x)=0.018$~eV; (III) $\Gamma_{ms}=0$, $t_{ma1}=1.6$~eV, $t_{ma2}=1.6$~eV,  $\gamma_{d0}=2.5\times10^{-4}$~eV, $\gamma_{0e}=1.5\times10^{-3}$~eV and $t_{p}(x)=0.018$~eV.
Other parameters producing the plots are: $\varepsilon_{a1}=0.16$~eV, $\varepsilon_{a2}=2.07$~eV, $\varepsilon_{h}=0.48$~eV, $\varepsilon_{l}=2.39$~eV, $t_{12}=1.5$~eV, $\Gamma_{ae}=0.8$~eV, $\omega_{0}=1.905$~eV, $\Delta_{p0}=0.003$~eV, $m_{0}=0.015$~eV, $m_{p}=0.015$~eV, $\gamma_{pe}=0.16$~eV, $\gamma_{dp}=0.016$~eV, $T=8$~K, $V_{st}=-2.5$~V.}.
In case I, a broadband emission in the STML spectra can be observed in Fig. \ref{Fig02} (I). This is from the radiative decay of the gap plasmon, 
while the molecular exciton does not participate to the transport. The other two cases are more interesting, which we focus in the following.

In case II, the interaction between the molecular exciton and the gap plasmon occurs, which results in coherent energy transfer between them. This interaction generates a sharp dip in the broadband emission spectra, as shown in Fig. \ref{Fig02} (II). The resulting asymmetric line-shape is a signature of the Fano resonance. 
Essentially, the single molecule only couples to the substrate in this case, no tunneling electrons excite the single molecule directly. But it can be excited by the gap plasmon indirectly. Also shown in the figure are the separate contributions of the flux from the gap plasmon and the molecular exciton. The spectrum of the molecular exciton is a normal Lorentzian-like peak, while that of the plasmon shows the typical Fano line shape and contributes dominantly to the total spectrum.
In case III,
the molecule is underneath the STM tip. Both optical fields can be excited directly by the tunneling electrons. Their coherent interaction
results in a Fano-shaped emission spectrum shown in Fig.~\ref{Fig02} (III). In this case, a sharp peak instead of dip is observed. The signature of the molecular exciton
becomes dominant, in contrast to case (II). 
In Appendix \ref{C}, through a simple model, we show that the asymmetric line shapes originate from the Fano interference between the two photon fields.

We now apply our theory to consider three recent experiments. We show that they fall into one of the above discussed three regimes. In the first experiment, the molecule is attached to the metallic electrodes (tip and substrate) through molecular linkers\cite{chong2016narrow}, corresponding to case III. In the other two, the molecule lies on thin insulating layer deposited on the metal substrate\cite{zhang2017sub,imada2017single}. The relative position of the tip and molecule can be adjusted to cover all the three regimes.

\subsection{A suspended molecular wire}
STM-induced narrow-line emission from a single molecular
emitter (H2P) connected to the tip and the substrate through oligothiophene linkers was reported in Ref.~ \onlinecite{chong2016narrow}.
The light spectra exhibits an asymmetric line shape in broad background, with the peak position
closely associated with the emission energy of the fused H2P molecule. 
The oligothiophene wires 
decouple the H2P emitter from the substrate and the tip. The length of the linker can be adjusted by lifting the STM tip away from the substrate.  Therefore, the distance between tip and substrate plays a key role in achieving molecular luminescence.
Here we simulate the evolution of tip-substrate distance by adjusting the non-radiative decay parameter $\gamma_{0e}$ of the photon field $\omega_0$ while keeping all other parameters fixed.

\begin{center}
\begin{figure}[h]
\includegraphics[scale=0.85]{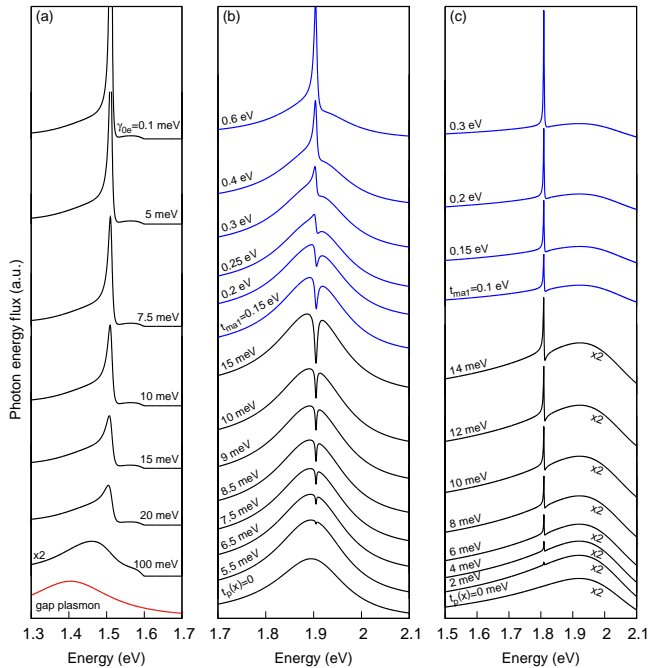}
\caption{(a) The photon energy flux for different values of the non-radiative damping $\gamma_{0e}$ with other parameters fixed. The red line represents the spectrum of gap plasmon for $V_{st}=-3$~V. Other parameters are: $\varepsilon_{a1}=0.16$~eV, $\varepsilon_{a2}=1.0$~eV, $\varepsilon_{h}=1.5$~eV, $\varepsilon_{l}=3.01$~eV, $t_{12}=0.2$~eV, $\Gamma_{ae}=0.2$~eV, $\Gamma_{ms}=0.0$, $t_{ma1}=0.2$~eV, $t_{ma2}=0.2$~eV, $\omega_{0}=1.51$~eV, $\omega_{p}=1.41$~eV, $m_{0}=0.01$~eV, $m_{p}=0.003$~eV, $t_{p}(x)=0.022$~eV, $\gamma_{pe}=0.25$~eV, $\gamma_{d0}=1\times10^{-5}$~eV, $\gamma_{dp}=0.008$~eV, $T=4.5$~K, $V_{st}=1.6$~V.
(b) The photon emission spectrum for different values of $t_p(x)$ without tip-molecule coupling (black lines), and for different values of tip-molecule coupling $t_{ma1}$ when $t_p(x)=15$ meV. Other parameters are the same with Fig.~\ref{Fig02} (II).
(c) Similar to (b), with different parameters $\omega_{0}=1.81$~eV, $\omega_{p}=2.0$~eV, 
$\varepsilon_{h}=0.48$~eV, $\varepsilon_{l}=2.29$~eV,
$\gamma_{0p}=0.35$~eV, $\gamma_{dp}=0.02$~eV, $T=4$~K, $\Gamma_{ms}=0.1$ and $V_{st}=-2.3$~V. The other parameters are the same as in Fig. \ref{Fig02} (II). }\label{Fig04}
\end{figure}
\end{center}

Figure~\ref{Fig04}(a) plots the emission spectrum (photon energy flux versus frequency/energy $\omega$) for several values of $\gamma_{0e}$ (proportional to the lifetime of the molecular excited state-LMES).  For the short distance case, i.e., the most part of the molecular linker is adsorbed on the substrate, it is difficult to observe a well-defined fluorescence from the molecular emitter because of the quenching of molecular luminescence, i.e., the
LMES is very short.
In this case, the spectrum observed is similar to most STM-induced
light emission experiments, showing a broad gap plasmon spectra 
[see $\gamma_{0e}=100$~meV in Fig.~\ref{Fig04}(a)].
With increased tip-substrate distance, the non-radiative decay becomes smaller. This results in an increased LMES. EL from the molecule can then be observed as a peak in the spectrum. The intensity of the peak becomes stronger with further decoupling from the substrate.
This is similar to the case (III) in Fig.~\ref{Fig02}.

\subsection{Single molecule on insulating layer}
In Refs.~\citenum{zhang2017sub} and \citenum{imada2017single},  STM-induced light emission from a single molecule decoupled from substrate by insulating NaCl layer was studied.
It was found that the relative tip-molecule position [$t_{p}(x)$] modifies the emission spectrum significantly. 
Figure~\ref{Fig04} (b) and (c) show the tip position-dependent flux obtained from our theory, corresponding results from Refs.~\cite{zhang2017sub,imada2017single}. In the case the molecule and the tip are far apart [$t_p(x)=0$], the two optical fields do not couple directly. Since the molecule does not participate the electron tunneling process, the molecular luminescence is not observed.  Approaching the tip to the molecule generates a nonzero $t_{p}(x)\neq0$, the coupling of the molecular exciton and gap plasmon mode opens the energy transfer channel between them. A sharp dip [(b)] or peak [(c)] develops due to the Fano interference. The interaction between the molecule exciton and the
gap plasmons can be tuned by varying the tip position near molecule. This allows one to control the hybridization of the two states. To provide a quantitative description that can be compared with the experimental data, we simulate the tip distance-dependent light spectra by adjusting the $t_{p}(x)$ ranging from $0$~to $15$~meV, all the main features of the experimental results are reproduced by our theory, e.g., the dip/peak structure becomes more and more prominent as $t_{p}(x)$ increases. 
We note that, although we consider only one exciton mode in Fig.~\ref{Fig04} (c), in the experiment, two peaks are observed corresponding to transition dipoles along the two ligands axes of H2Pc. The two dipoles are not degenerate due to the breaking of four-fold rotational symmetry in H2Pc.

Encouraged by the good agreement between our results and the experimental data, we go one step further. We study here the effect of the electronic coupling between the tip and the molecule on the spectrum.
The blue lines in Fig.~\ref{Fig04} (b) and (c) show the evolution of the spectrum with increasing electronic coupling while fixing the other parameters. We can see that the contribution from the molecular exciton becomes larger with stronger tip-molecule coupling. The  Fano dip in Fig.~\ref{Fig04} (b) gradually develops into an asymmetric peak, resembling the case in Fig.~\ref{Fig04} (a). These results show the importance of electronic subsystem on the photon emission spectrum. This has hitherto not been considered, and is beyond the simple model in the SI.
This prediction can be verified in experiment by changing the vertical tip-molecule distance, which has used in related studies\cite{schneider_optical_2010, schneider2012light}. 

\section{Summary}
In summary, we have developed a general theoretical model based on NEGF to investigate the single molecule-mediated light emission from a STM junction inspired by the recent experiments. Three different regimes are highlighted to explain the experimental results. 
Our model provides a clear description of the evolution of the spectra line shapes with the STM tip position. 
Moreover, this approach can also be used to study the light emission from other molecules such as DNA and RNA molecules, as the mismatch of base-pairs can be distinguished by the emission spectra\cite{park2002array,koenderink2015nanophotonics,vial2017single}. This provides a novel opportunity to detect the gene mutation. 

\begin{acknowledgements}
We are grateful to financial support from the National Natural Science Foundation of China (grant No.: 61371015).
\end{acknowledgements}


\appendix
\onecolumngrid
\setcounter{figure}{0}
\renewcommand{\thefigure}{C\arabic{figure}}
\section{Model and Hamiltonian}\label{A}
In our model as shown in Fig.~\ref{Fig01}(a), two extra `agents' from the substrate
($a_2$) and from the tip ($a_1$) are introduced, which couple to the substrate
and the STM tip, respectively. In reality, each of them is part
of the tip or the substrate. They are introduced mainly to avoid direct
coupling between the tip and substrate, which is convenient to apply the nonequilibrium Green's function (NEGF) 
theory. When the molecule is away from the tip, it couples
only to the substrate.  When it is underneath or very near the tip, it couples to the tip and the substrate through the two agents. 

The model Hamiltonian, consisting of the electronic reservoir, two photon fields and electron-photon interaction terms, can be defined as
\begin{equation}
\begin{split}
H=H_{el}+H_{ph}+H_{e-p},
\end{split}
\end{equation}
where
\begin{equation}
\begin{split}
H_{el}=H_b+H_a+H_m,
\end{split}
\end{equation}
\begin{equation}
\begin{split}
H_{ph}= H_{p0}+H_{p1}+H_{p01},
\end{split}
\end{equation}
\begin{equation}
\begin{split}
H_{e-p}= H_{ep0}+H_{ep1}.
\end{split}
\end{equation}
The Hamiltonian of the tip ($t$) and substrate ($s$) electrons is written as
\begin{equation}
H_b = \sum_{k\nu=t,s} \varepsilon_{k\nu}c^\dagger_{k\nu}c_{k\nu},
\end{equation}
where $c^\dagger_{k\nu}$ $(c_{k\nu})$ creates (annihilates) an
electron in the $\nu$ (tip or substrate) reservoir with momentum $k$ and energy $\varepsilon_{k\nu}$.
Hamiltonian of the agents, including coupling to $s$ and $t$, is
\begin{equation}
H_a = \sum_{i=1,2} \varepsilon_{a_i} d_i^\dagger d_i + (t_{12} d^\dagger_1 d_2+h.c.) + \sum_{kt}\left(t_{1t} d_1^\dagger c_{kt}+h.c.\right)+ \sum_{ks}\left(t_{2s} d_2^\dagger c_{ks}+h.c.\right),
\end{equation}
 where $d_i^\dagger$ $(d_i)$ creates (annihilates) an electron on the agent $i=1,2$ with energy $\varepsilon_{a_i}$, $t_{12}$ is the tunnel coupling between two agents,
 and $t_{1t}$ $(t_{2s})$ is the agent-reservoir electron transfer coupling. Hamiltonian of the molecule is
\begin{equation}
H_m = \sum_{i=h,l} \left[\varepsilon_i d_i^\dagger d_i + \sum_{j={1,2}}(t_{ij} d^\dagger_i d_j+h.c.)+ \sum_{ks}\left(t_{is}d_i^\dagger c_{ks}+h.c.\right)\right],
\end{equation}
where $d_i^\dagger$ $(d_i)$ creates (annihilates) an electron on the molecular orbital $i=h,l$ with energy $\varepsilon_i$,
$t_{ij}$ is the tunnel coupling between molecule and agent, and $t_{is}$ is the molecule-reservoir (substrate) electron transfer coupling.

The Hamiltonian for two photon fields are
\begin{equation}
H_{p0} = \hbar\omega_0 \left(\frac{1}{2}+a_0^\dagger a_0\right),
\end{equation}
\begin{equation}
H_{p1} = \hbar\omega_p \left(\frac{1}{2}+a_p^\dagger a_p\right) ,
\end{equation}
and
\begin{equation}
H_{p01} =  t_p(x) a^\dagger_p a_0 + h.c.,
\end{equation}
where $a_0^\dagger$ $(a_0)$ and $a_p^\dagger$ $(a_p)$ create (annihilate) photons in the two photon fields.
The term $H_{p01}$ is the coupling Hamiltonian between the two photon fields, and the coupling parameter $t_{p}(x)$ depends on the distance $x$
between the tip and the molecule.

The interaction between the photon (plasmon) mode with the electronic system is described within
the rotating wave approximation \cite{scully1999quantum}
\begin{equation}\label{11}
H_{ep0} = m_0 (d_h^\dagger d_l a_0^\dagger + d_l^\dagger d_h a_0),
\end{equation}
\begin{equation}\label{12}
H_{ep1} = m_p (d_{2}^\dagger d_{1} a_p^\dagger + d_{1}^\dagger d_{2} a_p),
\end{equation}
where $m_{0}$ and $m_{p}$ are the coupling parameters of molecular exciton-photon and  the agent-gap plasmon, respectively.

\section{The NEGF method}\label{B}
The NEGF method \cite{meir1992landauer,wang2008quantum,haug2008quantum,mahan2013many} is a powerful tool to investigate the
luminescence properties of the STM junction with consideration of the electron-photon coupling. We first define the photon Green's functions in the Keldysh contour with time on the contour as $\tau$
\begin{equation}
D_{ij}(\tau,\tau^{\prime}) = -\frac{i}{\hbar}\langle \mathcal{T}_t \{a_i(\tau) a_j^\dagger(\tau^{\prime})\} \rangle.
\end{equation}
In real time,  six different components of the Green's function can be defined as (set $\hbar=1$)
\begin{equation}
\begin{split}
&D_{ij}^{t}(t,t^{\prime})=-i\theta(t-t^{\prime})\langle a_{i}(t)a_{j}^{\dagger}(t^{\prime})\rangle-
i\theta(t^{\prime}-t)\langle a_{j}^{\dagger}(t^{\prime})a_{i}(t)\rangle,\\
&D_{ij}^{\bar{t}}(t,t^{\prime})=-i\theta(t-t^{\prime})\langle a_{j}^{\dagger}(t^{\prime})a_{i}(t)\rangle-i\theta(t^{\prime}-t)\langle a_{i}(t)a_{j}^{\dagger}(t^{\prime})\rangle,\\
&D_{ij}^{<}(t,t^{\prime})=-i\langle a_{j}^{\dagger}(t^{\prime})a_{i}(t)\rangle,\\
&D_{ij}^{>}(t,t^{\prime})=-i\langle a_{i}(t)a_{j}^{\dagger}(t^{\prime})\rangle,\\
&D_{ij}^{r}(t,t^{\prime})=-i\theta(t-t^{\prime})\langle [a_{i}(t),a_{j}^{\dagger}(t^{\prime})]\rangle,\\
&D_{ij}^{a}(t,t^{\prime})=i\theta(t^{\prime}-t)\langle [a_{i}(t),a_{j}^{\dagger}(t^{\prime})]\rangle.\\
\end{split}
\end{equation}
For our purpose, the most suitable functions are the
$D^{<,>}$ and $D^{r,a}$. In general, $D^{r}$ is linked to the response function and $D^{<,>}$ is related to
the light emission spectra, which can be obtained from the Dyson-Keldysh equations
\begin{equation}
\begin{split}
&D_{}^{r}(\omega)=D_{0}^{r}(\omega)+D_{0}^{r}(\omega)\Pi^{r}_{t}(\omega)D_{}^{r}(\omega),\\
&D_{}^{<}(\omega)=D_{}^{r}(\omega)\Pi^{<}_{t}(\omega)D_{}^{a}(\omega).
\end{split}
\end{equation}
Without electron-photon interaction, the Green's function $D_{0}^{r}$ for the bare photon system can be solved exactly using equation of motion method. $\Pi_{t}=\Pi_{el}+\Pi_{ev}+\Pi_{d}$ is the total photon self-energy, where
$\Pi_{el}$, $\Pi_{ev}$, and $\Pi_{d}$ account for the interaction with electrons, non-radiative decay, and radiative decay, respectively.
We consider wide-band environment and detector, such that 
$\Pi_{ev}$  and $\Pi_d$ can be expressed as
\begin{equation}
\begin{split}
&\Pi^{r}_{ev}(\omega)=-\frac{i}{2}{\rm diag}\{\gamma_{0e},\gamma_{pe}\},\\
&\Pi^{r}_{d}(\omega)=-\frac{i}{2}{\rm diag}\{\gamma_{d0},\gamma_{dp}\},\\
\end{split}
\end{equation}
and
\begin{equation}
\begin{split}
&\Pi^{<}_{ev,de}(\omega)=f^{b}(\omega)[\Pi^{r}_{ev,de}(\omega)-\Pi^{a}_{ev,de}(\omega)].
\end{split}
\end{equation}
Here, $\gamma_{0e}$ ($\gamma_{pe}$) and $\gamma_{d0}$ ($\gamma_{dp}$) are the radiative and non-radiative dissipation rate of molecular exciton (gap plasmon) due to coupling to the environment and the detector, respectively, 
$f^{b}(\omega)=[e^{\omega/k_{B}T}-1]^{-1}$ is the Bose-Einstein distribution function with the temperature $T$. Based on standard SCBA, the self-energies due to electron-photon interaction are given by
\begin{equation}
\begin{split}
&\Pi^{r}_{el,mn}(\omega)=-i \sum_{ijkl}M_{ij}^{m}\int\frac{d\varepsilon}{2\pi}[G_{li}^{r}(\varepsilon)G_{jk}^{<}(\varepsilon-\omega)+G_{li}^{<}(\varepsilon)G_{jk}^{a}(\varepsilon-\omega)]M_{kl}^{n},\\
&\Pi^{<}_{el,mn}(\omega)=-i \sum_{ijkl}M_{ij}^{m}\int\frac{d\varepsilon}{2\pi}G_{li}^{<}(\varepsilon)G_{jk}^{>}(\varepsilon-\omega)M_{kl}^{n}.\\
\end{split}
\end{equation}

Similarly, we can define the Green's function for electrons
\begin{equation}
G_{ij}(\tau,\tau^{\prime}) = -i\langle \mathcal{T}_t \{d_i(\tau) d_j^\dagger(\tau^{\prime})\} \rangle.
\end{equation}
In the energy space, the retarded and lesser Green's functions can be calculated from Dyson-Keldysh
equations
\begin{equation}
\begin{split}
&G^{r}(\varepsilon)=G^{r}_{0}(\varepsilon)+G^{r}_{0}(\varepsilon)\Sigma^{r}_{ep}(\varepsilon)G^{r}(\varepsilon),\\
&G^{<}(\varepsilon)=G^{r}(\varepsilon)[\Sigma^{<}_{ep}(\varepsilon)+\Sigma^{<}_{0}(\varepsilon)]G^{a}(\varepsilon),\\
\end{split}
\end{equation}
where $G_{0}$ is the Green's function for
electronic system without electron-photon interaction, and $\Sigma_{0}=\Sigma_{a-t}+\Sigma_{a-s}+\Sigma_{m-s}$ is the electronic self-energy describing the coupling to the tip ($\Sigma_{a-t}$) and substrate ($\Sigma_{a-s},\Sigma_{m-s}$),
respectively. Using the wide-band approximation for the tip and substrate electrodes, we have
\begin{equation}
\begin{split}
&\Sigma_{a-t}^{r}(\varepsilon)=-\frac{i}{2} {\rm diag}
\{\Gamma_{a1},0,0,0\},\\
&\Sigma_{a-s}^{r}(\varepsilon)=-\frac{i}{2}{\rm diag}\{0,0,0,\Gamma_{a2}\},\\
&\Sigma_{m-s}^{r}(\varepsilon)=-\frac{i}{2}{\rm diag}\{0,\Gamma_{ls},\Gamma_{hs},0\},
\end{split}
\end{equation}
and
\begin{equation}
\begin{split}
&\Sigma^{<}_{a-t}(\varepsilon)=-f^{e}_{t}(\varepsilon)[\Sigma^{r}_{a-t}(\varepsilon)-\Sigma^{a}_{a-t}(\varepsilon)],\\
&\Sigma^{<}_{a,m-s}(\varepsilon)=-f^{e}_{s}(\varepsilon)[\Sigma^{r}_{a,m-s}(\varepsilon)-\Sigma^{a}_{a,m-s}(\varepsilon)],\\
\end{split}
\end{equation}
where $\Gamma_{ai(=1,2)}$, and $\Gamma_{i(=l,h)s}$
are the linewidth functions, $f^{e}_{\nu}(\varepsilon)=[1+e^{(\varepsilon-\mu_{\nu})/k_{B}T}]^{-1}$ is the Fermi-Dirac distribution function for the electrode $\nu=t,s$
with the chemical potential $\mu_{\nu}$ and the temperature $T$,  $\mu_{s}-\mu_{t}=eV_{st}$ is the
tip-substrate voltage drop.
The self-energies ($\Sigma^{r,<}_{ep}$) due to electron-photon coupling are given within SCBA
\begin{equation}
\begin{split}
\Sigma^{r}_{ep,mn}(\varepsilon)&=-i\sum_{ijkl}M_{mn}^{i}D_{0,ij}^{r}(\omega=0)M_{kl}^{j}\int\frac{d\varepsilon}{2\pi} G_{lk}^{<}(\varepsilon)\\
&+i\sum_{ijkl}M_{mi}^{k}\int\frac{d\omega}{2\pi}[G_{ij}^{r}(\varepsilon-\omega)D_{kl}^{<}+G_{ij}^{<}(\varepsilon-\omega)D_{kl}^{r}+G_{ij}^{r}(\varepsilon-\omega)D_{kl}^{r}]M_{jn}^{l},\\
\Sigma^{<}_{ep,mn}(\varepsilon)&=i\sum_{ijkl}M_{mi}^{k}\int\frac{d\omega}{2\pi}G_{ij}^{<}(\varepsilon-\omega)D_{kl}^{<}(\omega)M_{jn}^{l}.\\
\end{split}
\end{equation}
The  interaction matrix $M$ describes molecule (agent)-photon field coupling, which can be divided into two types of contributions:
(a) $m_0 d_i^\dagger d_j a_0^\dagger$ $(i,j=h,l,i\neq j)$ in Eq. (\ref{11}) describe excitation and de-excitation between two molecular orbits, and (b) $m_p d_i^\dagger d_j a_p^\dagger$ $(i,j=1,2,i\neq j)$ in Eq. (\ref{12})
describe transitions between two electronic states of the agents that couple to the plasmon field.

Following the standard procedure\cite{haug2008quantum,mahan2013many,gao2016optical}, the energy flux of photon can be expressed as
\begin{equation}
\begin{split}
J_{ph}^{\alpha}=\int\frac{d\omega}{2\pi}\omega {\rm Tr}[\Pi^{<}_{\alpha}(\omega)D^{>}(\omega)-\Pi^{>}_{\alpha}(\omega)D^{<}(\omega)],\quad \alpha=el,ev,d.
\end{split}
\label{eq:mw}
\end{equation}
As expected, the conservation of energy $J_{ph}^{el}+J_{ph}^{ev}+J_{ph}^{d}=0$ is satisfied in steady state within SCBA.
For characterizing the luminescence properties of a STM junction, we may define the flux probed by the detector by let $\alpha=d$.

In the lowest order approximation to the electron-photon coupling, we can replacing $D^>$ and $D^<$ in Eq.~(\ref{eq:mw}) using $D^>_0$ and $D_0^<$. After the replacement, we can see that: (1) the energy spectrum goes into the photonic system is determined by both the electronic and the photonic system through $\Pi_e$ and $D_0$, respectively. The Fano effect is reflected in the photonic system $D_0$, especially the gap plasmons, as analyzed in Appendix~\ref{C}. 

\section{The formula for Fano resonance}\label{C}
To obtain a standard formula of Fano resonance, we consider a simple model to describe the coupling between the molecular exciton and the gap plasmon [see Fig. \ref{Fig01}(b)], in which the photon transport due to molecular exciton is regarded as a  scatter\cite{stefanski2003quantum,miroshnichenko2010fano}. Then the correction to the Green's function
for gap plasmon reads
\begin{equation}
\begin{split}
D_{p}^{r}=D_{p}^{0,r}+D_{p}^{0,r}\Pi_{p-m}^{r}D_{p}^{r}.
\end{split}
\end{equation}
$D_{p}^{r}$ $(D_{p}^{0,r})$ is the retarded Green's function of gap plasmon with (without) interaction, which can be expressed in the form of T-matrix by iterating
\begin{equation}\label{C2}
\begin{split}
D_{p}^{r}=D_{p}^{0,r}+D_{p}^{0,r}T_{p-m}D_{p}^{0,r},
\end{split}
\end{equation}
where
\begin{equation}
\begin{split}
T_{p-m}=\frac{\Pi_{p-m}^{r}}{1-\Pi_{p-m}^{r}D_{p}^{0,r}}, \Pi_{p-m}^{r}=\frac{T_{p-m}}{1+T_{p-m}D_{p}^{0,r}}.
\end{split}
\end{equation}
$\Pi_{p-m}^{r}$ represents retarded self-energy due to the coupling between gap plasmon and molecular exciton. For non-interacting photons, the expression for T-matrix
obtained by equation of motion takes the form
\begin{equation}
\begin{split}
T_{p-m}=t_{p}(x)D_{m}^{r}t_{p}^{\ast}(x),
\end{split}
\end{equation}
where $D_{m}^{r}$ is the retarded Green's function of the molecular exciton.
\begin{center}
\begin{figure}
\includegraphics[scale=0.15]{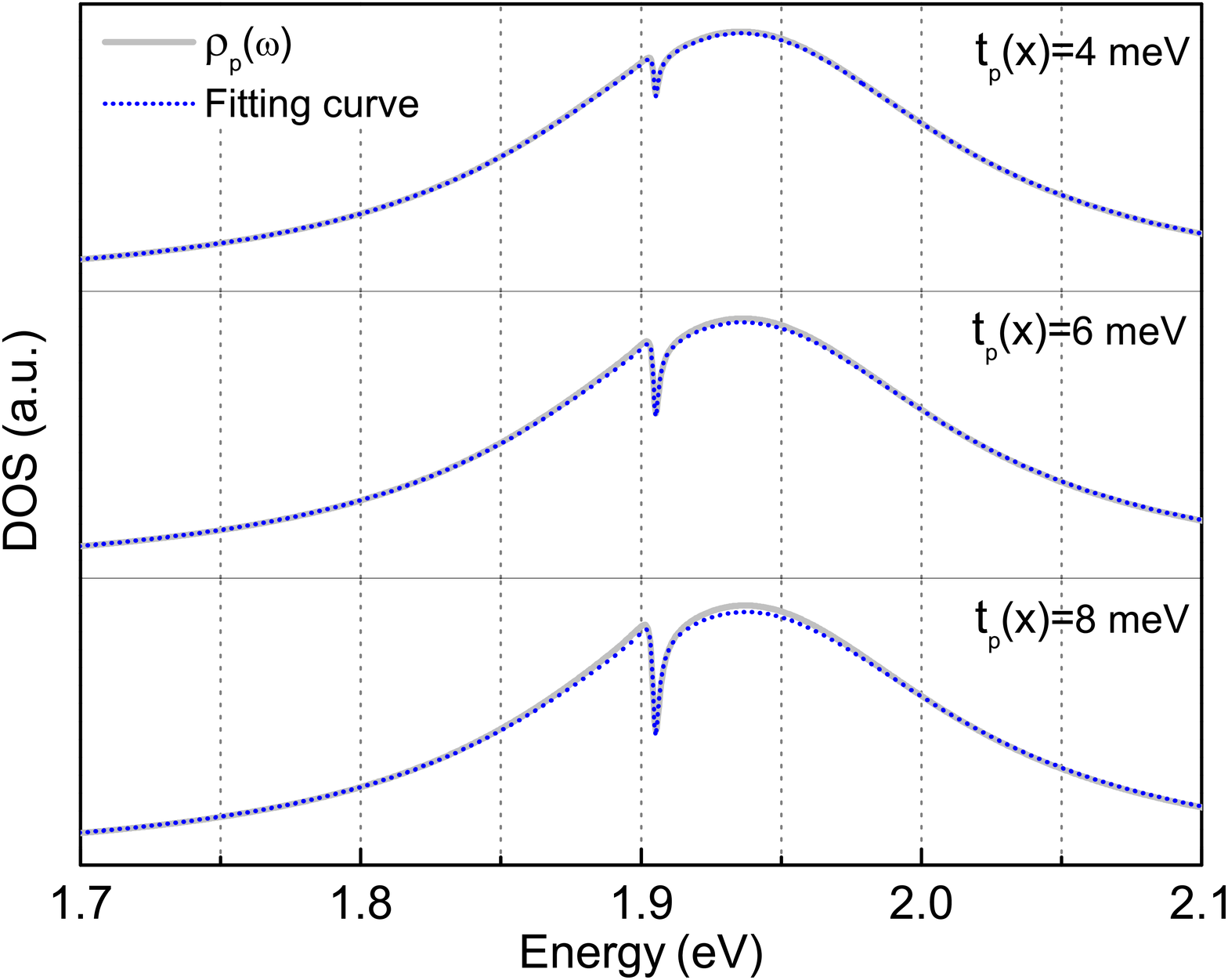}
\caption{The gray lines represent spectral
density $\rho_{p}(\omega)$ of gap plasmon with $\Delta_{p0}=0.03$~eV. The blue short dotted lines are the fitting curves for
the Fano line shape with Eq. (\ref{C11}). The fitting parameters [$\gamma_{t}$, $\omega_{t}$, $\gamma_{0}$, q] are
[0.0493, 1.905, -0.0009, 0.3397]~eV, [0.0935, 1.905, -0.001, 0.3411]~eV and [0.1367, 1.905, -0.0012, 0.3431]~eV for $t_{p}(x)$=4meV, 6meV and 8meV. The
other parameters are the same as in Fig.~\ref{Fig02}(II) in main text.}\label{FigA1}
\end{figure}
\end{center}
By using the definition of the spectral density of the gap plasmon with and without interaction: $\rho_{p}(\omega)=-\Im D_p^{r}(\omega)/\pi$ and
$\rho_{p}^{0}(\omega)=-\Im D_p^{0,r}(\omega)/\pi$, and taking the imaginary part of Eq. (\ref{C2})
\begin{equation}
\begin{split}
\rho_{p}(\omega)=\rho_{p}^{0}+\rho_{p}^{0}(\omega)\Im D_{p}^{0,r}(\omega)[\Im T_{p-m}(\omega)(q^{2}-1)-2q\Re T_{p-m}(\omega)],
\end{split}
\end{equation}
where we have defined 
\begin{equation}
\begin{split}
q=-\frac{\Re D_{p}^{0,r}(\omega)}{\Im D_{p}^{0,r}(\omega)}=\frac{\mathcal{Q}}{\gamma}.
\end{split}
\end{equation}
Here, $q$ is the Fano-factor, $\mathcal{Q}=t_{p}^{2}(x)\Re D_{p}^{0,r}$ and $\gamma=\pi t_{p}(x)^{2}\rho_{p}^{0}(\omega)$.
In the noninteracting case, the $D_{m}^{r}$ can be written as
\begin{equation}
\begin{split}
D_{m}^{r}=\frac{1}{\omega-\omega_{0}-\mathcal{Q}+i\gamma+i\gamma_{dr}}.
\end{split}
\end{equation}
$\gamma_{dr}$ represents the coupling with the electronic system, the environment and the detector. So, the $T_{p-m}$ takes the form
\begin{equation}
\begin{split}
T_{p-m}=\frac{t_{p}(x)t_{p}^{\ast}(x)}{\omega-\omega_{0}-\mathcal{Q}+i\gamma+i\gamma_{dr}}.
\end{split}
\end{equation}
We introduce the $\epsilon=(\omega-\omega_{0}-\mathcal{Q})/(\gamma+\gamma_{dr})$, then $T_{p-m}$ can be expressed as
\begin{equation}
\begin{split}
T_{p-m}&=t_{p}(x)t_{p}^{\ast}(x)\frac{\epsilon(\gamma+\gamma_{dr})-i(\gamma+\gamma_{dr})}
{\epsilon^{2}(\gamma+\gamma_{dr})^{2}+(\gamma+\gamma_{dr})^{2}}.\\
\end{split}
\end{equation}
Substituting the expression of $\Im D_{p}^{0,r}$ and $T_{p-m}$ into $\rho_{p}(\omega)$, we get
\begin{equation}\label{C11}
\begin{split}
\rho_{p}(\omega)
&=\rho_{p}^{0}(\omega)+\gamma_{t}[\rho_{p}^{0}(\omega)]^{2}\frac{q^{2}+2q\epsilon-1}
{\epsilon^{2}+1},\\
\end{split}
\end{equation}
with $\gamma_{t}=\pi t_{p}(x)t_{p}^{\ast}(x)/\gamma_{0}$, $\gamma_{0}=\gamma+\gamma_{dr}$, $\epsilon=(\omega-\omega_{t})/\gamma_{0}$ and $\omega_{t}=\omega_{0}+\mathcal{Q}$. Here, $(q^{2}+2q\epsilon-1)
/(\epsilon^{2}+1)$ indicates the spectral density of the gap plasmon has a Fano profile determined by the parameter $q$. Note that $\gamma_{t}$ characterizes the effective coupling strength of the gap plasmon with the molecular exciton.

To demonstrate the Fano resonance is an universal phenomenon in the single molecule-based STML experiments, we can fit the numerical results of the spectral density of gap plasmon with the formula in Eq. (\ref{C11}). The results of the fitting are shown in Fig. \ref{FigA1} (see the blue dotted lines). It is found that the Fano resonance can well describe the gap plasmon spectral density, which implies that the asymmetric line shape with a dip in the light spectra originates from the Fano resonance between the molecular exciton and gap plasmon with different lifetimes. Moreover, the Fano line shape becomes prominent as the $t_{p}(x)$ increases as well as the  corresponding fitting parameter $\gamma_{t}$. Therefore, one can use the Eq. (\ref{C11}) to fit and predict the light emission spectra in the single molecule-based STML experiments.
On the other hand, Eq. (\ref{C11}) can be expressed as a more frequently used form to fit the experimental result
\begin{equation}
\begin{split}
\rho_{p}(\omega)
&=\rho_{p}^{0}(\omega)F(\epsilon),\\
\end{split}
\label{C12}
\end{equation}
where $F(\epsilon)=(\epsilon+q)^{2}/(\epsilon^{2}+1)$ is Fano function. 





%


\twocolumngrid
\bibliography{REF}

\end{document}